\documentclass[11pt]{article}
\usepackage[affil-it]{authblk}
\usepackage[usenames,dvipsnames]{xcolor}
\usepackage{amsfonts}
\usepackage{amsmath,amsthm,amssymb,amsfonts}
\usepackage{enumerate}
\usepackage[english]{babel}
\usepackage{graphicx}	
\usepackage{subcaption}
\usepackage[margin=0.99in]{geometry}
\usepackage{url}
\usepackage{todonotes}
\usepackage{bbm}
\usepackage{tikz}
\usetikzlibrary{chains}
\usetikzlibrary{fit}
\usepackage{pgflibraryarrows}		
\usepackage{pgflibrarysnakes}		
\usepackage{authblk}
\usepackage{makecell}
\usepackage{multirow}
\usepackage{epsfig}
\usetikzlibrary{shapes.symbols,patterns} 
\usepackage{pgfplots}

\usepackage{hyperref}
\hypersetup{colorlinks=true,citecolor=blue,linkcolor=blue,filecolor=blue,urlcolor=blue,breaklinks=true}

\usepackage{nicefrac}
\usepackage{mathtools}
\usepackage{booktabs}

\usepackage{pgf, tikz}
\usetikzlibrary{arrows, automata}

\usepackage{enumitem}
\usepackage[authoryear]{natbib} 

\theoremstyle{plain}

\newtheorem*{theorem*}{Theorem}
\newtheorem*{proposition*}{Proposition}

\theoremstyle{definition}

\newtheorem*{definition*}{Definition}

\newcommand*{\RR}{\mathbb{R}}

\DeclarePairedDelimiterX{\inner}[2]{\langle}{\rangle}{#1, #2}

\title{A Mixture of Experts Vision Transformer for High-Fidelity Surface Code Decoding}

\author[1]{Nguyen Hoang Viet}
\author[1]{Nguyen Manh Hung}
\author[1]{Hoang Ta}
\author[2]{Van Khu Vu}
\author[3]{Yeow Meng Chee}
\affil[1]{ \small Hanoi University of Science and Technology, Vietnam}
\affil[2]{VinUniversity, Vietnam}
\affil[3]{Singapore University of Technology and Design, Singapore}
\date{}

\begin{document}
\maketitle
\begin{abstract}
    Quantum error correction is a key ingredient for large scale quantum computation, protecting logical information from physical noise by encoding it into many physical qubits. Topological stabilizer codes are particularly appealing due to their geometric locality and practical relevance. In these codes, stabilizer measurements yield a syndrome that must be decoded into a recovery operation, making decoding a central bottleneck for scalable real time operation. Existing decoders are commonly classified into two categories. Classical algorithmic decoders provide strong and well established baselines, but may incur substantial computational overhead at large code distances or under stringent latency constraints. Machine learning based decoders offer fast GPU inference and flexible function approximation, yet many approaches do not explicitly exploit the lattice geometry and local structure of topological codes, which can limit performance. In this work, we propose QuantumSMoE, a quantum vision transformer based decoder that incorporates code structure through plus shaped embeddings and adaptive masking to capture local interactions and lattice connectivity, and improves scalability via a mixture of experts layer with a novel auxiliary loss. Experiments on the toric code demonstrate that QuantumSMoE outperforms state-of-the-art machine learning decoders as well as widely used classical baselines.
\end{abstract}
            
\section{Introduction}
\label{sec:introduction}
A central goal of quantum information research is to build a large quantum computer that exploits quantum mechanics to solve problems beyond the reach of classical computers. The main challenge is the fragility of quantum information. Unlike classical data, quantum states tend to decohere quickly when they interact with the environment~\cite{nielsen2010quantum}. Finding the right balance between protecting qubits from decoherence and still allowing controlled access for computation has proven extremely challenging~\cite{preskill2018quantum}. 

A natural approach to protecting quantum information is to encode each logical qubit into a larger register of physical qubits using quantum error correction. Quantum error correcting codes (QEC) enable reliable storage and processing by extracting \emph{syndrome} information through measurements of suitable observables that diagnose errors while preserving the encoded quantum state. Early constructions of QEC include Shor's code~\cite{shor1995scheme} and Steane's code~\cite{steane1996multiple}. Subsequently, inspired by the viewpoint of classical linear coding theory, where a code is specified by parity checks and the syndrome is interpreted as the outcome of these constraints, one arrives at the \emph{stabilizer formalism}~\cite{gottesman1997stabilizer}. In this formalism, the codespace is specified by a commuting set of Pauli constraints, and the corresponding measurements yield a syndrome used for decoding and recovery without revealing the logical information.

The stabilizer formalism encompasses several important code families, including Calderbank--Shor--Steane (CSS) codes~\cite{calderbank1996good}, quantum low density parity check codes~\cite{mackay2004sparse,panteleev2022asymptotically}, and topological stabilizer codes. Among these, topological stabilizer codes, such as the surface code~\cite{kitaev1997quantum,kitaev2003fault,fowler2012surface, terhal2015quantum}, are widely viewed as a promising route toward scalable fault tolerant quantum computation, since their stabilizer checks are compatible with local operations on a two dimensional layout and they admit comparatively high accuracy thresholds~\cite{raussendorf2007fault,fowler2012surface}. Recent experiments have further provided evidence that increasing the code distance can suppress logical errors, including demonstrations of surface code scaling and below threshold operation~\cite{google2023suppressing,google2025quantum}.


A central practical challenge in implementing topological stabilizer codes is \emph{decoding}: given a measured syndrome, one seeks to infer a likely physical error pattern, or more generally the induced logical effect, in order to apply an appropriate recovery. Existing decoding methods are often grouped into two broad categories: classical algorithmic decoders and machine learning based decoders. Two widely used classical decoders for topological stabilizer codes, such as the surface code, are Minimum Weight Perfect Matching (MWPM)~\cite{dennis2002topological,higgott2022pymatching} and union find decoders~\cite{huang2020fault}. While these approaches are effective and come with strong algorithmic guarantees in important regimes, their runtime and implementation overhead can still be significant at the scale required for large code distances and real time operation. More recently, machine learning based (ML-based) decoders have emerged as a promising alternative, benefiting from fast GPU inference and the ability to learn complex correlations between qubits and noise~\cite{bausch2024learning}. 
Nevertheless, state-of-the-art existing learning based approaches \cite{wang2023transformer, park2025hierarchical, choukroun2024deep} typically leverage only the parity-check matrix, and therefore do not yet fully exploit key structural properties of topological codes, such as geometric locality, translational symmetries, and the way errors propagate through stabilizer constraints, which are crucial for achieving fast, robust, and reliable decoding at scale. This motivates the design of decoders that combine the efficiency of modern learning methods with inductive biases tailored to the underlying code structure.

The geometric characteristics of topological codes described above naturally correspond to the notion of spatial inductive bias in vision models. Such architectures exploit spatial correlations among groups of pixels to learn latent image representations. Consequently, \emph{Vision Transformer} (ViT) based architectures emerge as promising candidates for decoding topological quantum codes, as their attention mechanisms enable the modeling of both local and global information. Moreover, there is an increasing demand for models that enhance representational capacity without compromising inference efficiency, which is essential for real-time decoding. In this regard, \emph{Mixture of Experts} (MoE) models provide an attractive approach to scaling model capacity while maintaining efficient inference through sparse, conditional computation. MoE architectures have become a cornerstone of the success of numerous large-scale foundation language models, including Mixtral 8x7B~\cite{jiang2024mixtral}, Gemini 2.5~\cite{comanici2025gemini}, and DeepSeek V3~\cite{liu2024deepseek}. However, integrating MoE architectures is far from a straightforward task due to load balancing challenges and instabilities caused by discontinuous routing mechanisms.
Despite their potential, standard MoE models are difficult to implement effectively because of routing discontinuities and uneven workload distribution among experts. Instead of direct expert routing, SoftMoE \cite{puigcerver2023sparse} maps tokens into aggregated slots, thereby removing the discontinuities typical of sparse MoE models. This design is ideal for quantum topological decoders, which aim to identify and group error patterns to prevent logical errors from compromising the system.

Motivated by these considerations, we propose QuantumSMoE, a decoder that utilizes the ViT architecture and SoftMoE component to achieve superior accuracy for surface code decoding (given the extracted syndrome). By discretizing the surface code into a series of patch embeddings through a novel convolution layer, our framework effectively captures intrinsic geometric properties of the surface codes. It further incorporates a specialized adaptive masking mechanism designed to enforce spatial locality, ensuring each patch selectively attends only to its most pertinent neighboring regions.
Additionally, a novel auxiliary loss is implemented to optimize the token-to-slot assignment within the SoftMoE blocks, fostering more effective grouping of similar inputs. To the best of our knowledge, this is among the first demonstrations that Mixture-of-Experts can be used to improve the performance of ML-based decoders for quantum error-correcting codes. While our methodology is applicable to a wide range of surface code variants, we present results for the toric code, as it serves as a standard periodic benchmark for controlled evaluation.

In summary, QuantumSMoE makes three main contributions as follows.
\begin{itemize}
    \item \textbf{A novel ViT based decoder} for toric codes that explicitly incorporates geometric and topological characteristics into the model architecture.
    \item \textbf{A slot orthogonality loss} designed to encourage the SoftMoE module to more accurately assign similar patch inputs to the same subset of experts.
    \item \textbf{Extensive numerical evaluations} across a range of physical error rates and code distances demonstrate that the proposed approach yields superior logical error rate reduction compared to classical baselines, including MWPM, MWPM-Corr~\cite{fowler2013optimal}, and BP LSD~\cite{hillmann2025localized}, as well as the state-of-the-art ML-based decoder QECCT~\cite{choukroun2024deep}.
\end{itemize}

\section{Related Work}
\label{sec:related_work}

Building on the decoding challenges discussed above, we briefly review representative classical and machine learning based decoding methods for topological stabilizer codes and related families.

\textbf{Classical Decoders.} For topological stabilizer codes, a long line of work has studied classical decoding algorithms that trade off accuracy and computational efficiency. In general, computing the optimal recovery, for example via maximum likelihood decoding, is NP hard~\cite{kuo2020hardnesses}, which motivates the development of efficient approximation methods. Among classical decoders, tensor network methods~\cite{berezutskii2025tensor} are often regarded as some of the most accurate approaches for quantum topological codes. Nevertheless, their substantial computational cost and poor scalability with increasing code distance render them impractical for real world deployment. The Union Find (UF) decoder~\cite{huang2020fault} was developed to achieve near linear time complexity, making it particularly suitable for real time hardware decoding. A more accurate and widely used alternative is Minimum Weight Perfect Matching (MWPM), which runs in polynomial time and can attain near optimal error thresholds under independent noise models~\cite{higgott2025sparse}. However, the standard MWPM formulation is not well suited for complex stabilizer errors, since it treats $X$ and $Z$ errors independently, even though a single qubit may experience both types of errors via $Y$ errors. To address this limitation, MWPM-Corr was introduced in~\cite{fowler2013optimal}, which improves decoding accuracy by accounting for correlations induced by $Y$ errors. For quantum low density parity check (QLDPC) codes, the BP-OSD decoder, which combines belief propagation (BP) with ordered statistics decoding (OSD), was proposed in~\cite{roffe2020decoding} as a powerful approach. More recently, BP-LSD~\cite{hillmann2025localized} has emerged as a leading decoder for QLDPC codes. Since the toric code is a prototypical instance of the LDPC family, it is particularly well suited to BP-LSD, which combines belief propagation (BP) with localized statistics decoding (LSD) to achieve state-of-the-art performance with significantly reduced computational overhead.

\textbf{Machine learning based Decoders.}
Machine learning based approaches have emerged as a promising direction for quantum error correction, as they enable fast inference and can learn complex interactions between physical qubits and stabilizers through deep neural architectures. A wide range of machine learning models have been proposed for scalable and accurate decoding, including multilayer perceptrons~\cite{chamberland2018deep,wagner2019symmetries}, long short term memory networks~\cite{baireuther2018machine}, convolutional neural networks~\cite{breuckmann2018scalable}, and transformer based architectures~\cite{choukroun2024deep,wang2023transformer,park2025hierarchical}. Recently, Google's AlphaQubit integrated a machine learning based decoder into a realistic superconducting qubit platform~\cite{bausch2024learning}, demonstrating the practical advantages of learning based decoders over classical approaches. Moreover, a range of hardware accelerators have been explored to further improve the execution speed of machine learning based decoders~\cite{zhang2020sparch,wang2021spatten,wang2025cim}. However, existing models do not explicitly exploit the geometric structure of the surface code, forcing the networks to learn redundant spatial relationships and making accurate inference under correlated errors more challenging. In contrast, our proposed QuantumSMoE explicitly embeds geometric information through a patch based representation and an adaptive masking mechanism that reflects stabilizer locality. Moreover, by integrating a Mixture of Experts architecture, our decoder can more accurately classify complicated errors at each decoding stage while incurring only modest computational overhead compared to non MoE based models.

\section{Preliminaries}
In this section, we review the necessary background on quantum error correcting codes, with an emphasis on the toric code. We also provide the basic background on the mixture of experts model.

\subsection{Quantum Error Correction}
\label{sec:quantum_code}

Quantum error correction aims to protect encoded (logical) quantum information from physical noise acting on a larger register of physical qubits. Because an unknown quantum state cannot be cloned and because direct measurements generally disturb the state, quantum error correcting codes are designed so that error information can be extracted \emph{indirectly} without revealing, and thus collapsing, the logical state. Concretely, one encodes a logical state into a subspace (the \emph{codespace}) $\mathcal{C}\subset (\mathbb{C}^2)^{\otimes n}$ such that typical errors map $\mathcal{C}$ into distinguishable error subspaces, enabling one to diagnose errors via suitable measurements and then apply a recovery operation.

\paragraph{Stabilizer codes and Pauli errors.}
A central and widely used class of quantum error correcting codes is given by \emph{stabilizer codes}, specified by an abelian subgroup
$S\subset \mathcal{P}_n$ of the $n$ qubit Pauli group $\mathcal{P}_n=\{\pm 1,\pm i\}\cdot \{I,X,Y,Z\}^{\otimes n}$, with $-I\notin S$.
The codespace is the simultaneous $+1$ eigenspace of all stabilizers:
\[
\mathcal{C} \;=\; \bigl\{\,|\psi\rangle : \forall g\in S,\; g|\psi\rangle = |\psi\rangle \,\bigr\}.
\]
Noise is often discretized as \emph{Pauli errors} $E\in\{I,X,Y,Z\}^{\otimes n}$ (up to an irrelevant global phase), where $X$ acts as a bit flip, $Z$ as a phase flip, and $Y=iXZ$ combines both.
Measuring a generating set $S=\langle g_1,\dots,g_m\rangle$ yields a binary \emph{syndrome} $s\in\mathbb{F}_2^m$, where
$s_j=1$ if and only if $E$ anticommutes with $g_j$ (that is, $g_jE=-Eg_j$), and $s_j=0$ otherwise.
This extracts error information without directly measuring the logical state.

Using the binary symplectic representation, a Pauli operator (up to phase) is encoded by a length $2n$ vector
$\varepsilon=(\mathbf{z}\mid\mathbf{x})\in\mathbb{F}_2^{2n}$ indicating the locations of $Z$ and $X$ components, and each stabilizer generator $g_j$ corresponds to a row vector
$h_j=(\mathbf{z}_j\mid\mathbf{x}_j)$.
Stacking these rows gives the \emph{parity check} (syndrome) matrix
$H\in\mathbb{F}_2^{m\times 2n}$, and the syndrome satisfies the relation
\[
s \;=\; H \cdot \varepsilon \quad (\text{over } \mathbb{F}_2),
\]
which encodes commutation and anticommutation constraints row by row (equivalently via the symplectic inner product).
Operators commuting with all stabilizers form the normalizer $N(S)$; elements of $N(S)\setminus S$ implement nontrivial \emph{logical operators}.
Given $s$, the decoding task is to infer a likely error representative $\hat\varepsilon$ (up to stabilizers) or to directly infer the induced logical operator $l = \mathbb{L}\hat{\varepsilon}$ represented by $\mathbb{L} \in \mathbb{F}_2^{2k \times 2n}$. This task is generally hard for exact maximum likelihood decoding and thus motivates approximate and learning-based decoders.

\subsection{Toric Code}
\label{sec:toric_code}

A particularly important stabilizer code family is the \emph{toric code}, which provides a prototypical example of a two dimensional topological stabilizer code. It is defined on a $L\times L$ square lattice with periodic boundary conditions, topologically a torus, with physical qubits placed on edges. We write $e$ for an edge of the lattice, $\delta(v)$ for the set of edges incident to a vertex $v$, and $\partial p$ for the set of boundary edges of a plaquette (face) $p$. The stabilizer generators are local Pauli operators: \emph{star} checks at vertices,
\[
A_v=\prod_{e\in \delta(v)} X_e,
\]
and \emph{plaquette} checks on faces,
\[
B_p=\prod_{e\in \partial p} Z_e.
\]
The codespace is the simultaneous $+1$ eigenspace of all $\{A_v,B_p\}$. The code encodes $k=2$ logical qubits and has distance $d$. Under Pauli noise, an $X$ type error flips the outcomes of neighboring $Z$ checks, producing pairs of plaquette defects, while a $Z$ type error flips neighboring $X$ checks, producing pairs of star defects. Hence, the syndrome can be interpreted geometrically in terms of localized defects on the lattice.

\begin{figure}[h!]
        \centering
        \includegraphics[width=\linewidth]{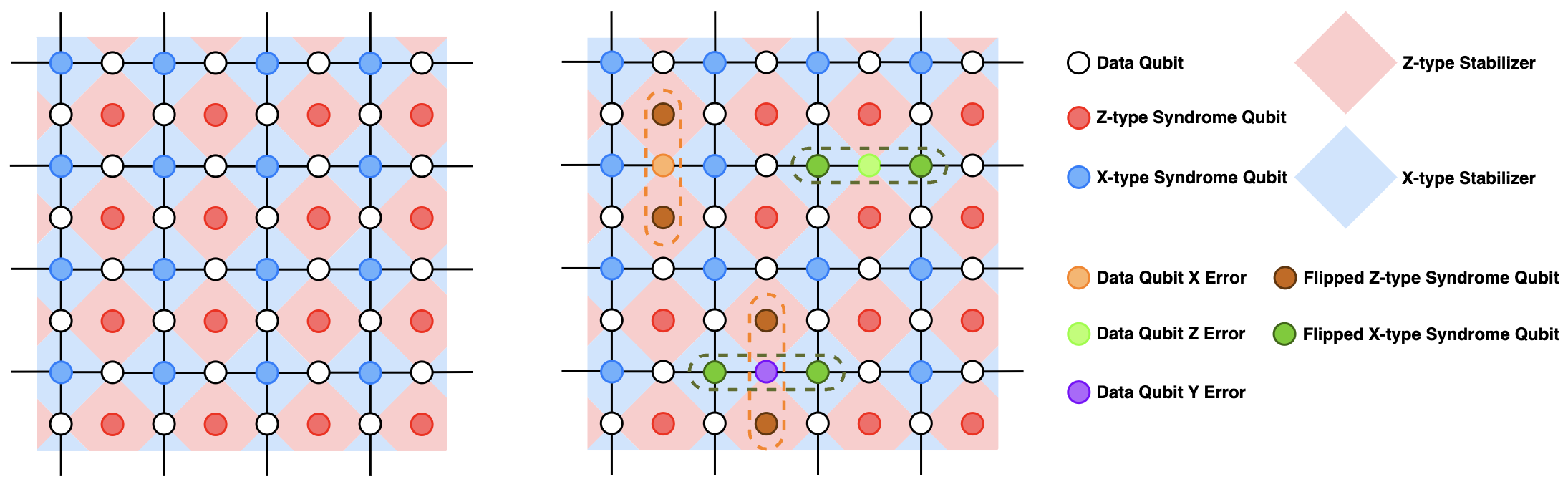}
        \caption{Error syndromes in the Toric code. $Z$-type syndrome qubits detect $X$ errors (yellow), while $X$-type syndrome qubits detect $Z$ errors (green); both types respond to $Y$ errors (purple). A single-qubit error typically flips the measurement outcomes of two adjacent syndrome qubits, complicating the decoding process as the error density increases. }
        \label{fig:toriccode}
    \end{figure}

\subsection{Mixture of Experts}
First introduced in~\cite{jacobs1991adaptive}, the Mixture of Experts (MoE) framework is an ensemble architecture that combines multiple specialized submodels through an adaptive gating mechanism.
In a typical MoE layer, let there be $n$ expert $f_1,\ldots,f_n$, and a sequence of input tokens represented by $\mathbf{X} = [x_1,\ldots,x_N] \in \RR^{N \times d}$, where $N$ is the number of tokens in the sequence. Consequently, the output of the dense MoE layer can be reformulated as 
\begin{equation}
\mathcal{F}_{\text{dense}}\left(\mathbf{X};\mathbf{\Theta}, \{\mathbf{W}_i\}_{i=1}^n\right) = \sum_{i=1}^n\mathcal{G}(\mathbf{X};\mathbf{\Theta})_if_i(\mathbf{X};\mathbf{W}_i),
\end{equation}
where $\mathcal{G}(\mathbf{X};\mathbf{\Theta})_i := \text{softmax}(g(\mathbf{X};\mathbf{\Theta}))_i = \dfrac{\exp(g(\mathbf{X};\mathbf{\Theta})_i)}{\sum_{j=1}^n\exp(g(\mathbf{X};\mathbf{\Theta})_j)}$ represents the gating score used to route the input to each expert $f_i$. To improve computational efficiency, the sparse MoE~\cite{shazeer2017outrageously} is implemented, which limits expert activation to $k < n$ per input token. This is achieved through a specialized gating function as follows:
\begin{equation}
    \mathcal{G}(\mathbf{X};\mathbf{\Theta})_i := \text{softmax}(\text{TopK}(g(\mathbf{X};\mathbf{\Theta}), k))_i,
\end{equation}
where $\text{TopK}(g(\mathbf{X};\mathbf{\Theta}),k)_i = g(\mathbf{X};\mathbf{\Theta})_i$ if $g(\mathbf{X};\mathbf{\Theta})_i$ is in the top-$k$ elements, and $0$ otherwise. To maintain low overhead, $k$ is generally assigned a value of 1 or 2. Consequently, the processing cost of the sparse MoE is only marginally higher than that of a basic MLP (Multi-Layer Perceptron) architecture. Despite its efficiency, the traditional sparse MoE architecture is often plagued by load imbalance, where a small subset of experts is overactive while others remain underutilized. Furthermore, the discrete nature of the top-$k$ gating mechanism can introduce training instabilities, complicating model convergence. 

Alternatively, a variant of sparse MoE, namely SoftMoE, are introduced without suffering these limitations. In particular, each expert $f_i$ processes $p$ slots, and the slots' gate is parameterized by $\mathbf{\Phi} \in \RR^{d \times (n\cdot p)}$.  The input slots $\widetilde{\mathbf{X}} \in \mathbb{R}^{(n\cdot p)\times d}$ are the result of combination of all the $n$ input tokens, defined by $\widetilde{\mathbf{X}} = \mathbf{D}^\top \mathbf{X}$, where 
$$
\mathbf{D}_{ij} = \dfrac{\exp((\mathbf{X}\mathbf{\Phi})_{ij})}{\sum_{i'=1}^N\exp((\mathbf{X}\mathbf{\Phi})_{i'j})}, \forall\, 1 \le i \le N, 1 \le j \le np.
$$
Then, the output of the softMoE is defined by $\mathcal{F}_{\text{SoftMoE}}(\mathbf{X};\mathbf{\Theta}, \{\mathbf{W}_i\}_{i=1}^n) = \mathbf{C}(\mathbf{X};\mathbf{\Theta})\widetilde{\mathbf{Y}}(\widetilde{\mathbf{X}};\{\mathbf{W}_i\}_{i=1}^n)$, where
$$\mathbf{C}(\mathbf{X};\mathbf{\Phi})_{ij} = \dfrac{\exp((\mathbf{X}\mathbf{\Phi})_{ij})}{\sum_{j'=1}^{np}\exp((\mathbf{X}\mathbf{\Phi})_{ij'})}, \forall\, 1 \le i \le N, 1 \le j \le np, $$
and 
$ \widetilde{\mathbf{Y}}(\widetilde{\mathbf{X}};\{\mathbf{W}_i\}_{i=1}^n)_j := f_{\lfloor j/p \rfloor}(\widetilde{\mathbf{X}}_j;\mathbf{W}_{\lfloor j/p \rfloor}),\forall \, 1 \le j \le np.$ 
Consequently, the computational overhead of the SoftMoE block is comparable to that of the original sparse MoE, as the total number of tokens routed to experts remains $n\cdot p = \mathcal{O}(N)$, consistent with standard sparse MoE architectures. 

\section{Proposed Method}
\label{sec:method}
\subsection{Framework Overview}
\begin{figure}
    \centering\includegraphics[width=0.8\linewidth, scale=0.8]{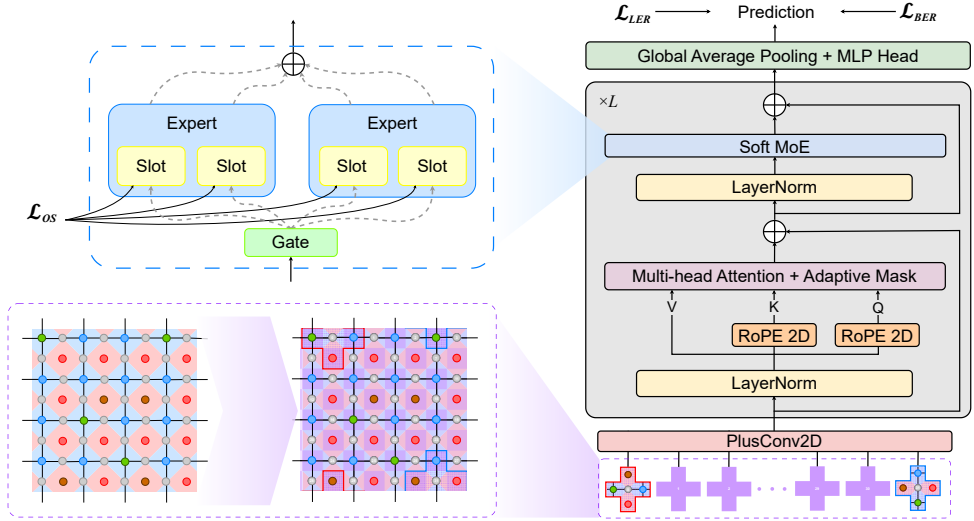}
    \caption{Overall framework of the proposed model. The qubit lattice of the toric code is first partitioned into patches and embedded using the proposed PlusConv2D convolutional layer. Within each Transformer block, the standard MLP is replaced by a SoftMoE layer, where discrepancy among slot representations is further promoted through the introduced auxiliary loss $\mathcal{L}_{\text{os}}$.}
    \label{fig:framework} 
\end{figure}
To improve decoding performance, we developed QuantumSMoE, a unified approach that incorporates the geometric properties of toric codes. Figure \ref{fig:framework} provides a visual overview of the proposed methodology.  Our methodology adopts the ViT backbone \cite{dosovitskiy2020image}, initiating the process by partitioning the qubit surface into discrete patches. These patches undergo embedding via PlusConv2D, a convolutional layer designed to capture local syndrome information effectively. To facilitate scale-invariance across code distances, we introduce Adaptive Masking within each Transformer block. This mechanism acts on the computed attention scores and is specifically tailored to the structural properties of toric codes. Upon combining the results of the attention module and the residual path, the data is fed into a SoftMoE block. To further enhance diversity and reduce similarity between the slot inputs routed to various experts, we implement a novel auxiliary loss, $\mathcal{L}_{\text{os}}$, derived from layer-wise slot representations. Finally, the token representations undergo global average pooling followed by an MLP head to predict the final error. 
\subsection{QuantumSMoE}

\hspace{0.5cm}\textbf{PlusConv2D.} We describe how spatial information from toric codes is incorporated into the model architecture. In toric codes, errors occurring on a data qubit induce flips in nearby syndrome qubits depending on the error type ($X$, $Y$, or $Z$), as illustrated in Fig.~\ref{fig:toriccode}. Consequently, effectively capturing such local interactions is crucial for modeling the relationships between data qubits and their neighboring syndrome qubits. Motivated by this observation, instead of employing a conventional $3\times3$ convolution that aggregates information from unrelated qubits, we design a specialized convolutional operator that includes only the four relevant neighboring syndrome qubits in the toric code. This operator has a plus-shaped receptive field and is referred to as PlusConv2D.

\begin{figure*}
\centering
\begin{subfigure}[t]{0.48\textwidth}
    \centering
    \includegraphics[scale=0.3]{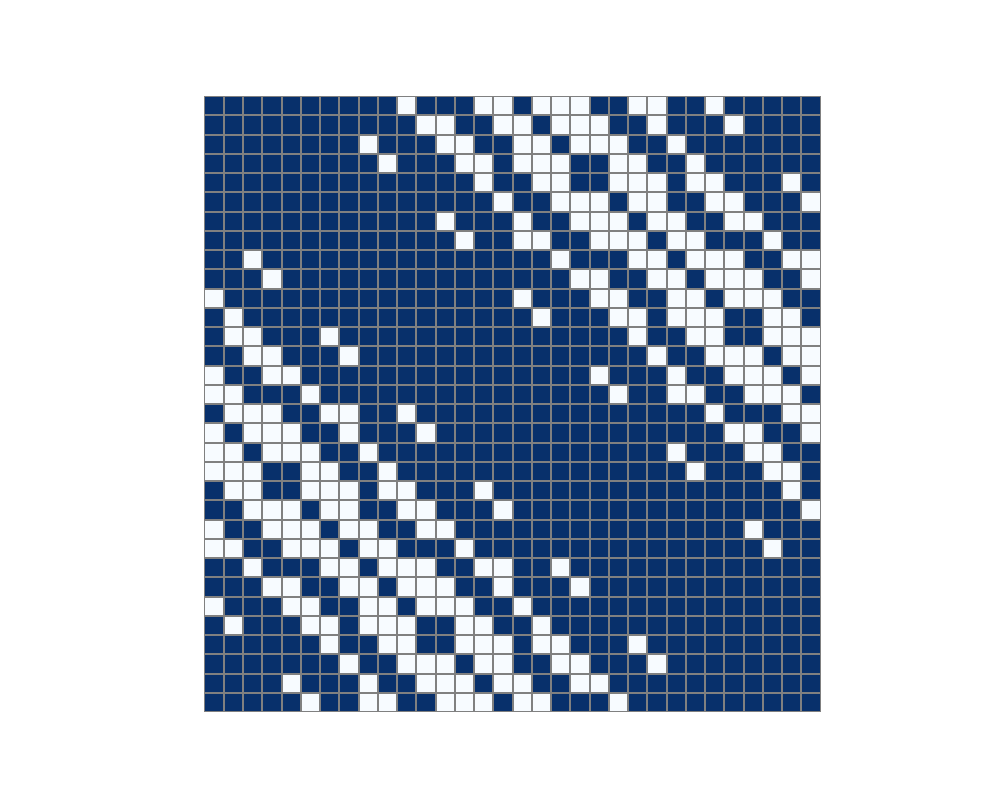}
    \caption{$L=4$}
\end{subfigure}
\hfill\begin{subfigure}[t]{0.48\textwidth}
    \centering
    \includegraphics[scale=0.3]{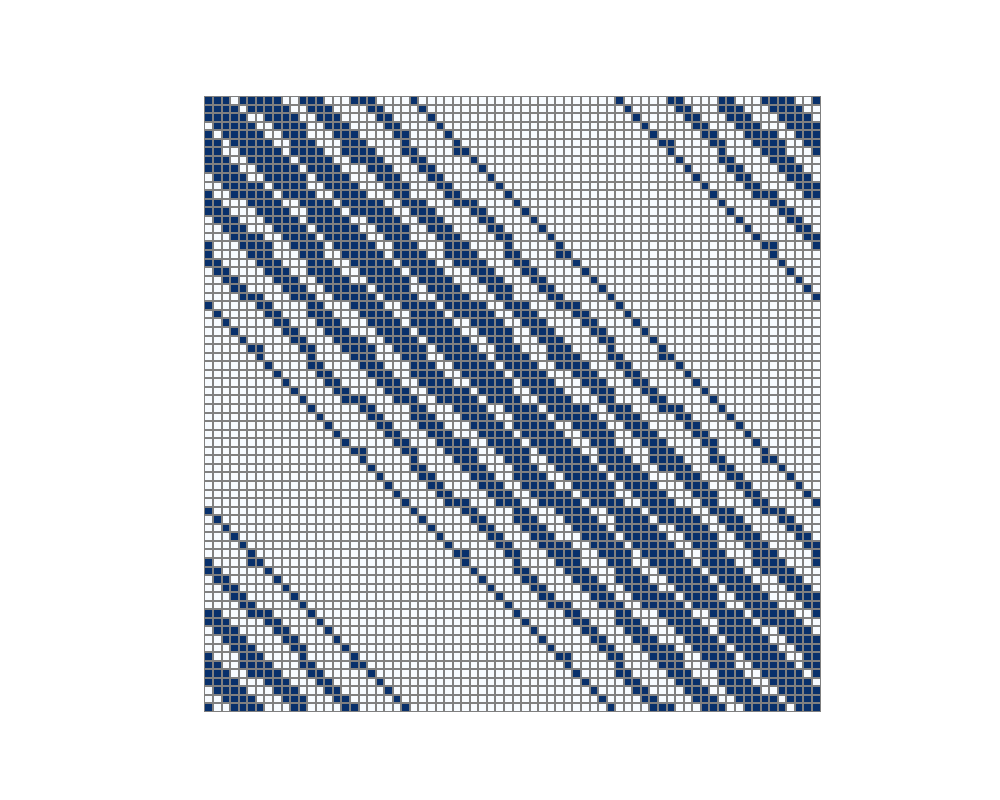}
\caption{$L=6$}
\end{subfigure}
\caption{Proposed adaptive masks at code distances $L=4$ and $L=6$.}
\label{fig:adaptive_mask}
\end{figure*}

\textbf{Adaptive Masking.} The proposed adaptive mask for code distances $L=4$ and $L=6$ is illustrated in Fig.~\ref{fig:adaptive_mask}. 
One of the central challenges in decoding surface codes, such as the toric code, lies in inferring the sequence of error locations from the observed syndrome measurements. In particular, when errors occur on pairs of qubits, the intermediate syndrome qubits may flip and subsequently return to their original states, making the error pattern ambiguous. To address this challenge, we model the intrinsic propagation of errors between qubits by imposing a structured mask on the attention scores. Specifically, two patches are allowed to attend to each other if and only if they share a common syndrome qubit, thereby capturing the neighboring connectivity dictated by the structural layout of the topological code. 

\textbf{Slot Orthogonal Loss Function.} Within each SoftMoE layer, we implement a novel auxiliary loss designed to encourage greater discrepancy among slot representations dispatched to different experts. The loss is defined as
\begin{equation}
    \mathcal{L}_{\text{os}} := \dfrac{1}{2(n-1)p}\sum_{\substack{1 \le i,j \le n\cdot p,\\ \lfloor i/p\rfloor \ne \lfloor j/p \rfloor
    }} \dfrac{\widetilde{\mathbf{X}}_i^\top \widetilde{\mathbf{X}_j}}{\|\widetilde{\mathbf{X}}_i\|\cdot \|\widetilde{\mathbf{X}}_j\|},
\end{equation}
where $n$ denotes the number of experts, $p$ is the number of slots per expert, and $\widetilde{\mathbf{X}}_i$ represents the embedding of the $i$-th slot. Minimizing $\mathcal{L}_{\text{os}}$ increases the dissimilarity between slot representations associated with different experts, thereby encouraging each expert to focus on more specialized inputs. Consequently, this loss also guides the gating mechanism to route input tokens to their most relevant experts, enabling the model to produce more specialized outputs tailored to specific types of quantum errors.

\textbf{Overall Loss Function.} The training objective of our model consists of a weighted combination of two primary loss terms, namely $\mathcal{L}_{\text{BER}}$ and $\mathcal{L}_{\text{LER}}$. Let $(s,\varepsilon)$ denote the observed syndrome and the corresponding physical error pattern, respectively. The bit-level error loss $\mathcal{L}_{\text{BER}}$ is defined as
\begin{equation}
\mathcal{L}_{\text{BER}} := \mathrm{BCE}(\mathrm{QuantumSMoE}(s), \varepsilon),
\end{equation}
where $\mathrm{BCE}$ denotes the binary cross-entropy loss. This term encourages the model to produce accurate predictions at the individual qubit level. To further enforce correctness at the logical operator level, we define the logical error loss $\mathcal{L}_{\text{LER}}$ as
\begin{equation}
\mathcal{L}_{\text{LER}} := \mathrm{BCE}(\mathbb{L}\mathrm{QuantumSMoE}(s), \mathbb{L}\varepsilon),
\end{equation}
where $\mathbb{L}$ represents the logical operator associated with the toric code. Hence, the overall training objective is then given by
\begin{equation}
\mathcal{L}_{\text{overall}} := \lambda_{\text{BER}}\cdot\mathcal{L}_{\text{BER}} + \lambda_{\text{LER}}\cdot\mathcal{L}_{\text{LER}} + \lambda_{\text{os}}\cdot\mathcal{L}_{\text{os}},
\end{equation}
where the weighting coefficients $\lambda_{\text{BER}}$, $\lambda_{\text{LER}}$ are selected by $0.5$ and $1$ as the setting in \cite{choukroun2024deep}, while $\lambda_{\text{os}}$ is set to $0.1$ via grid search to achieve optimal performance.


 \section{Experiments}
 \label{sec:experiment}
\subsection{Experimental Setting}
\begin{figure*}
\centering
\begin{subfigure}[t]{0.31\textwidth}
    \centering
    \includegraphics[scale=0.25]{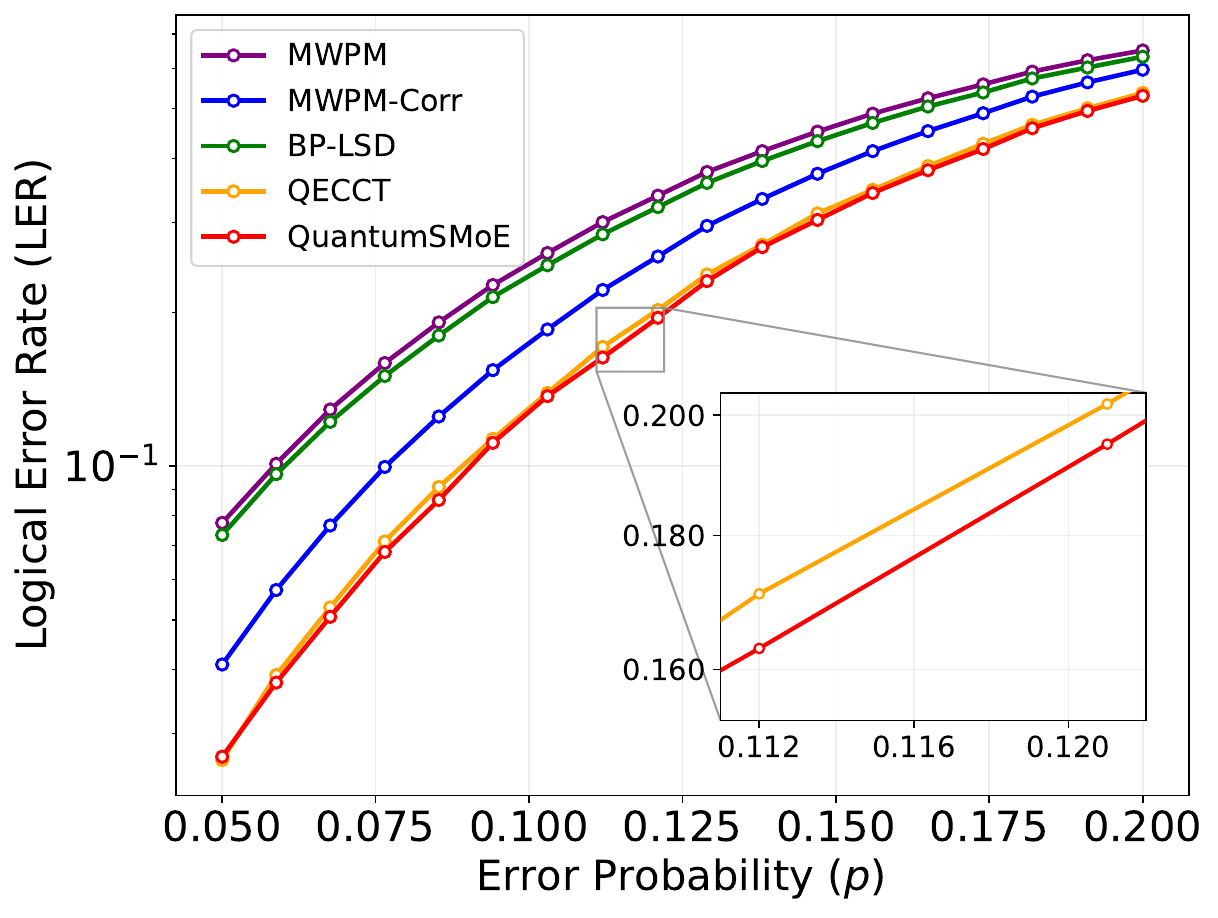}
    \caption{LER Comparison ($L=4$)}\label{fig:LER_L=4}
\end{subfigure}
\hfill\begin{subfigure}[t]{0.31\textwidth}
    \centering
    \includegraphics[scale=0.25]{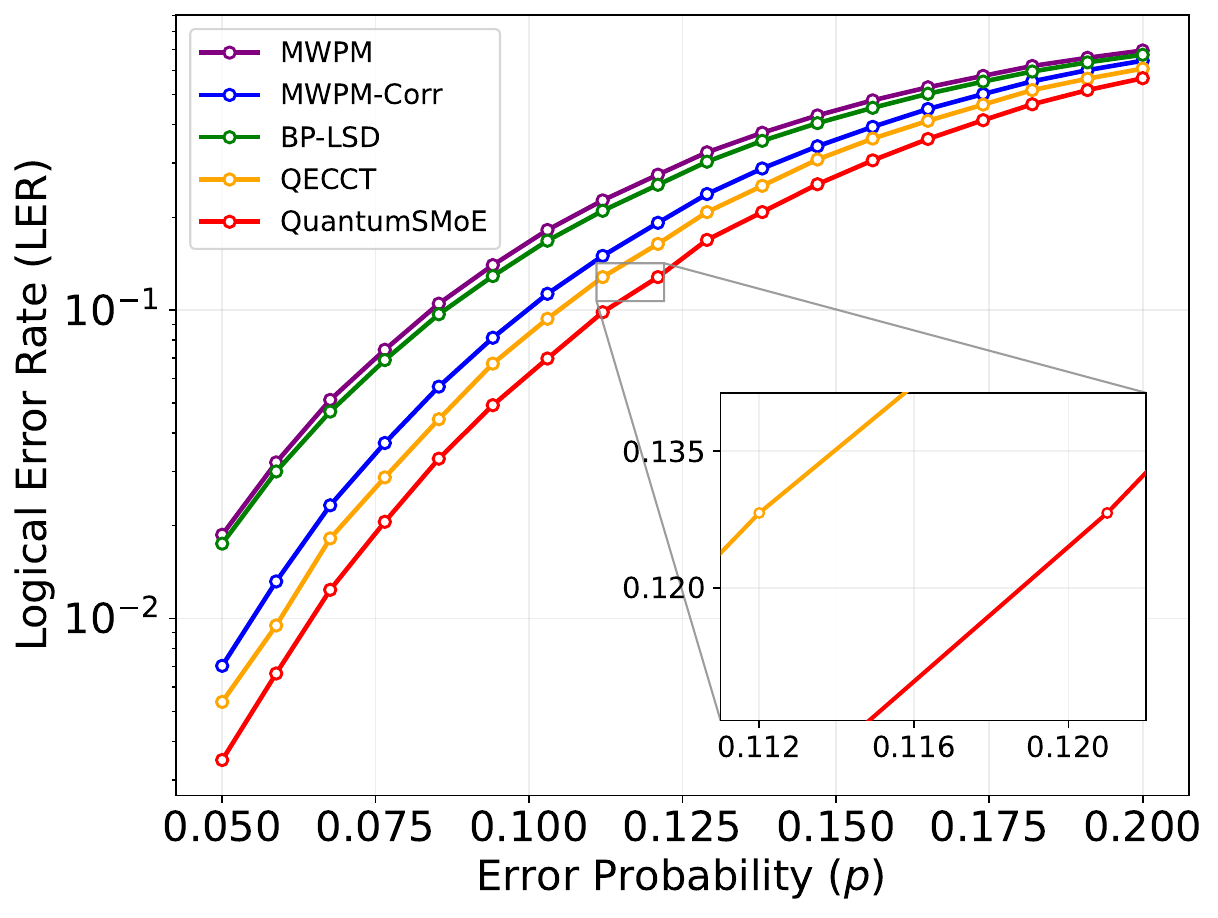}
\caption{LER Comparison ($L=6$)}\label{fig:LER_L=6}
\end{subfigure}
\hfill\begin{subfigure}[t]{0.31\textwidth}
    \centering
    \includegraphics[scale=0.25]{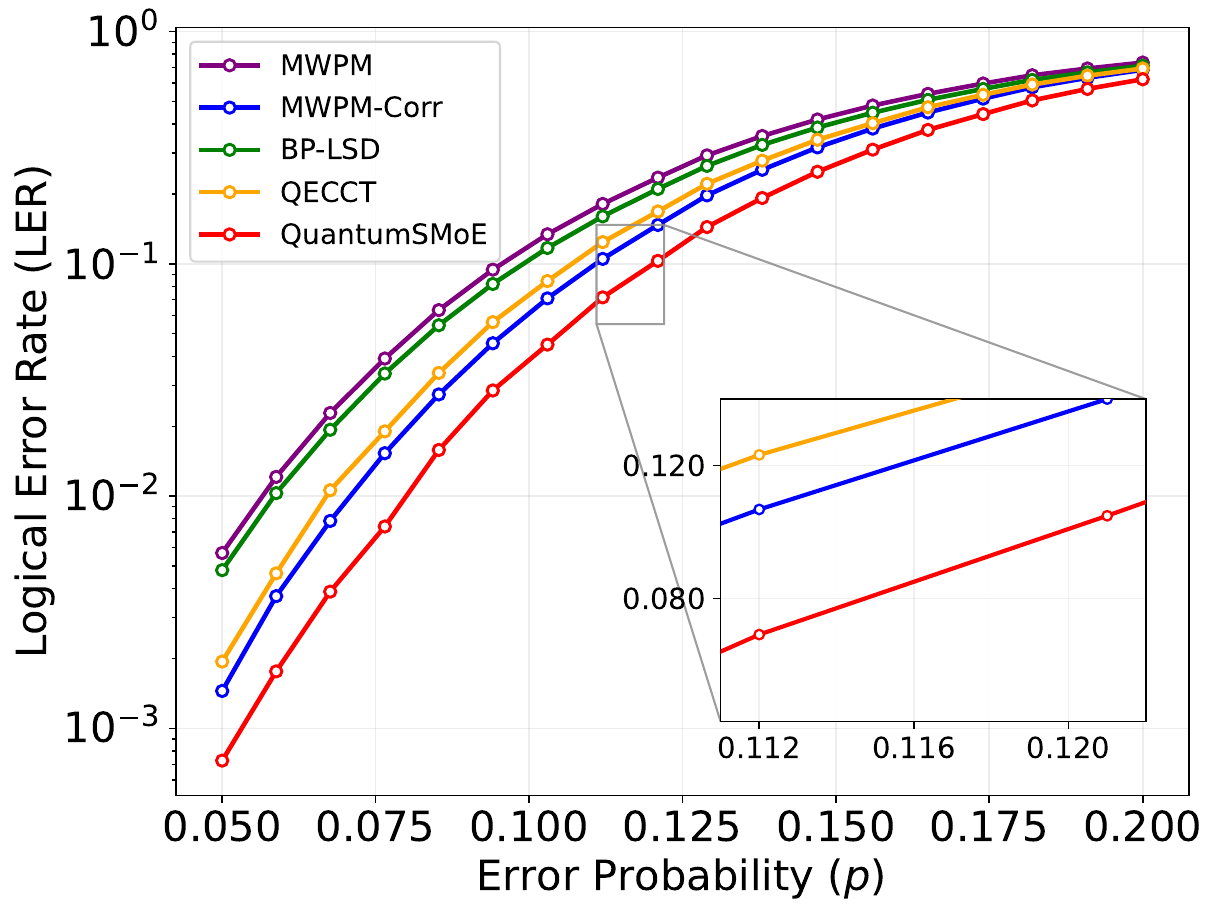}
\caption{LER Comparison ($L=8$)}\label{fig:LER_L=8}
\end{subfigure} \\
\begin{subfigure}[t]{0.31\textwidth}
    \centering
    \includegraphics[scale=0.25]{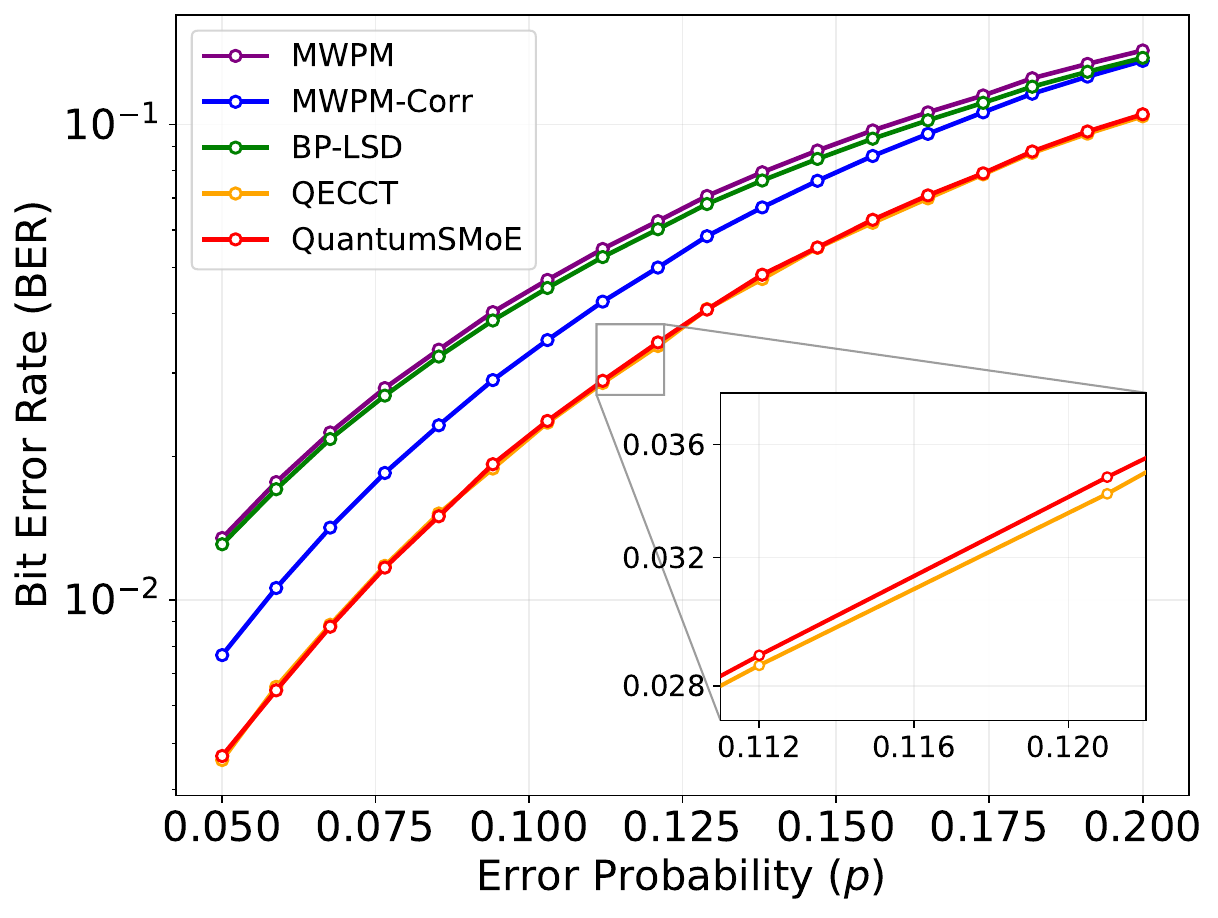}
    \caption{BER Comparison ($L=4$)}\label{fig:BER_L=4}
\end{subfigure}
\hfill\begin{subfigure}[t]{0.31\textwidth}
    \centering
    \includegraphics[scale=0.25]{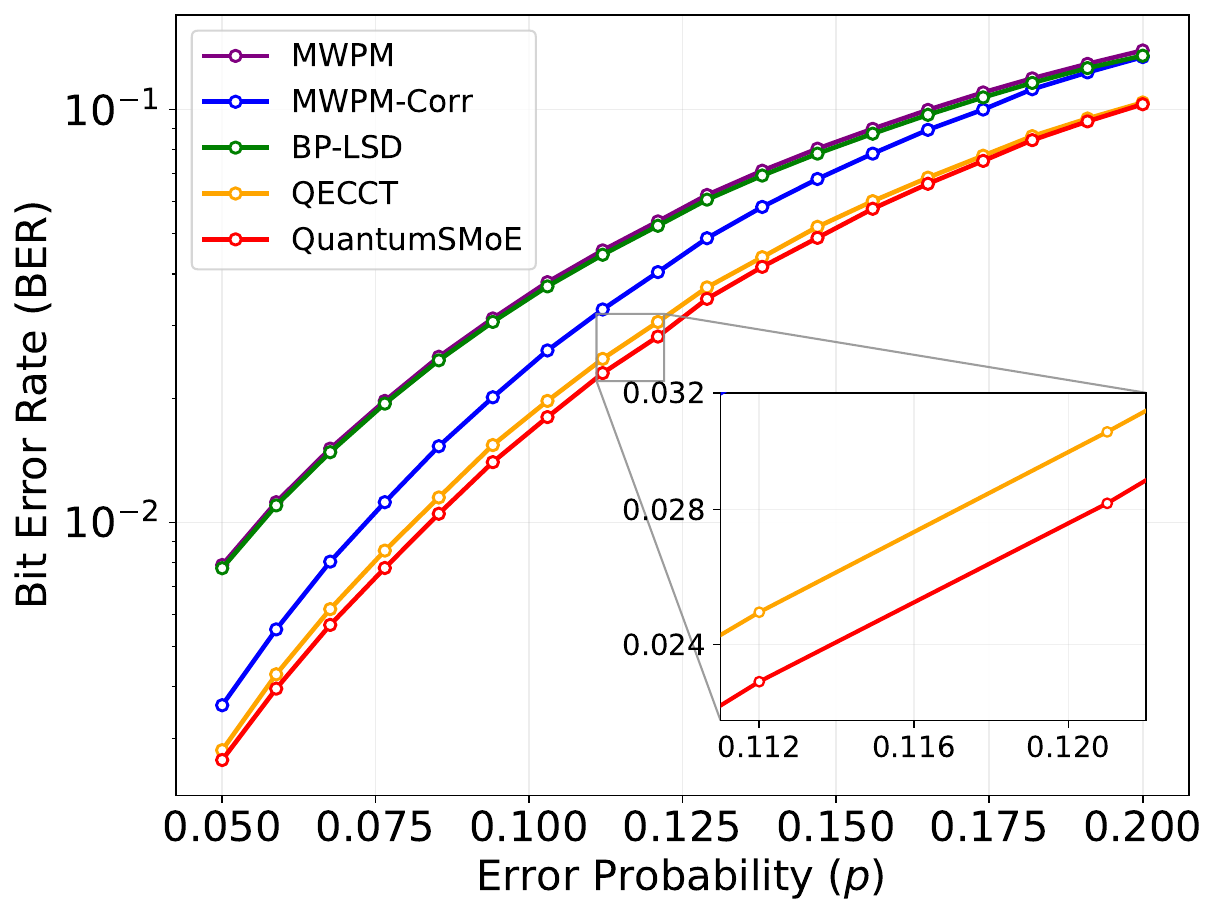}
\caption{BER Comparison ($L=6$)}\label{fig:BER_L=6}
\end{subfigure}
\hfill\begin{subfigure}[t]{0.31\textwidth}
    \centering
    \includegraphics[scale=0.25]{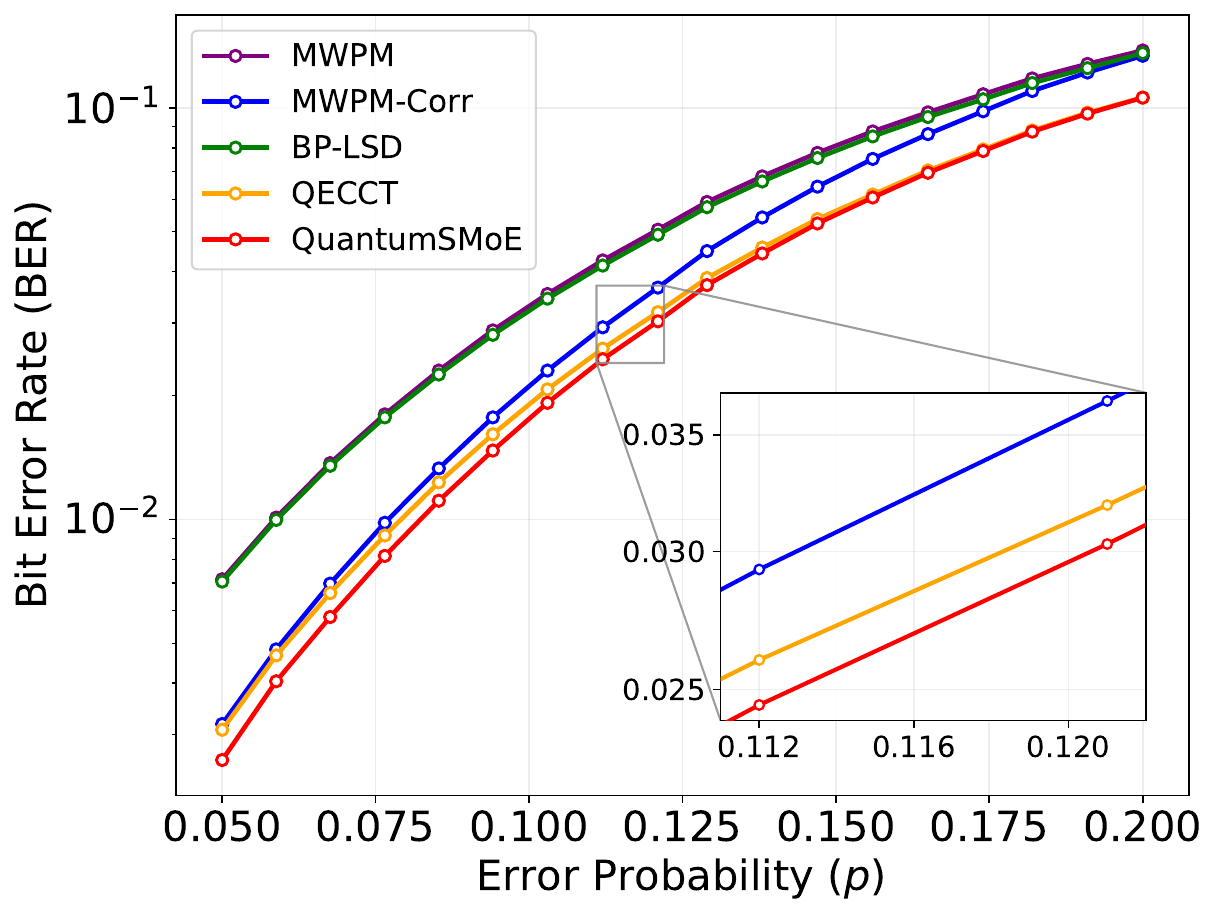}
\caption{BER Comparison ($L=8$)}\label{fig:BER_L=8}
\end{subfigure}
\caption{Comparison of Bit Error Rate (BER) and Logical Error Rate (LER) between our method and classical baselines, including MWPM, MWPM-Corr, BP-LSD, as well as the ML-based model QECCT. The results are evaluated over 18 distinct physical error rates ranging from $0.05$ to $0.2$. Our model consistently achieves lower LER than both classical and machine learning–based decoders.}
\label{fig:ler_ber}
\end{figure*}
\hspace{0.5cm}\textbf{Benchmark.} In our experimental setup, we evaluate the proposed QuantumSMoE framework on the toric code with code distances $L=4,6,8$. A depolarizing noise model is employed to generate physical errors and the corresponding correlated syndromes. Under this noise model, each Pauli error occurs with equal probability $\mathbb{P}(X) = \mathbb{P}(Y) = \mathbb{P}(Z) = p/3$, where the Pauli operator 
$Y$ is defined as $Y = iXZ$. We select 9 different values of the error rate 
$p$ in the range $[0.05,0.2]$ to generate the training data, and 18 values within the same range to construct the test data for evaluation. The implementation of the toric code is adopted from \cite{choukroun2024deep}.

\textbf{Baseline.} For the classical baselines, we first consider the widely used minimum-weight perfect matching (MWPM) decoder, which has a computational complexity of $\mathcal{O}[(n^3+n^2)\log(n)]$ and is implemented using the PyMatching package \cite{higgott2022pymatching}. This decoder treats $X$ and $Z$ errors independently. To capture correlated noise arising from $Y$ errors, we additionally evaluate an enhanced variant, MWPM-Corr \cite{fowler2013optimal}, which incorporates error correlations while preserving the same computational complexity. Furthermore, we include BP-LSD \cite{hillmann2025localized}, a state-of-the-art decoder for quantum low-density parity-check codes. For machine learning–based decoders, QECCT \cite{choukroun2024deep} is selected as the comparative baseline. 

\textbf{Metrics.} We report both the Logical Error Rate (LER) and the Bit Error Rate (BER). Specifically, BER measures the probability of incorrectly predicting a physical qubit error, while LER quantifies the probability of applying an incorrect logical operator after decoding based on the predicted errors.

\textbf{Implementation Details.} Our model is trained using 6 layers, with MoE layers applied in the final two blocks to reduce decoding latency. We set the embedding dimension to 128, following the same design adopted in QECCT. The SoftMoE layer is configured with 8 experts, where each expert processes 4 slots using a feedforward network with a hidden dimension of 512. The number of attention heads is set to $8$. For training, we use $5,760,000$ samples generated from nine physical error rates ranging from 0.05 to 0.2. The model is trained for 200 epochs using the AdamW optimizer, with an initial learning rate of 0.001 and a cosine decay schedule down to $10^{-6}$. Training is performed on a single NVIDIA RTX 4090 GPU with an AMD EPYC 7282 CPU (16 cores). 
\subsection{Main Results}

Figure~\ref{fig:ler_ber} presents a comparison of the proposed QuantumSMoE decoder with several leading practical decoders, including MWPM, MWPM-Corr, BP-LSD, and state-of-the-art machine learning–based approaches. For a code distance of $L=4$, our model achieves a slightly lower bit error rate than QECCT while attaining a superior LER. These observations suggest that QuantumSMoE more effectively captures correlated error patterns by exploiting the periodic connectivity and geometric structure of the Toric code, which are not explicitly incorporated in QECCT, thereby better prevents logical failures. Although QECCT consistently attains a lower BER, its logical error rate is higher than that of MWPM-Corr when $L=8$. This suggests that QECCT prioritizes minimizing per-qubit errors instead of capturing the global topological correlations necessary to prevent logical failures. Conversely, our architecture outperforms competing decoders in both BER and LER metrics across all configurations. These results demonstrate QuantumSMoE's proficiency in identifying complex error patterns at the individual qubit level and confirms its superior scalability to larger lattice sizes.

\begin{figure}
    \centering
    \includegraphics[width=0.5\linewidth]{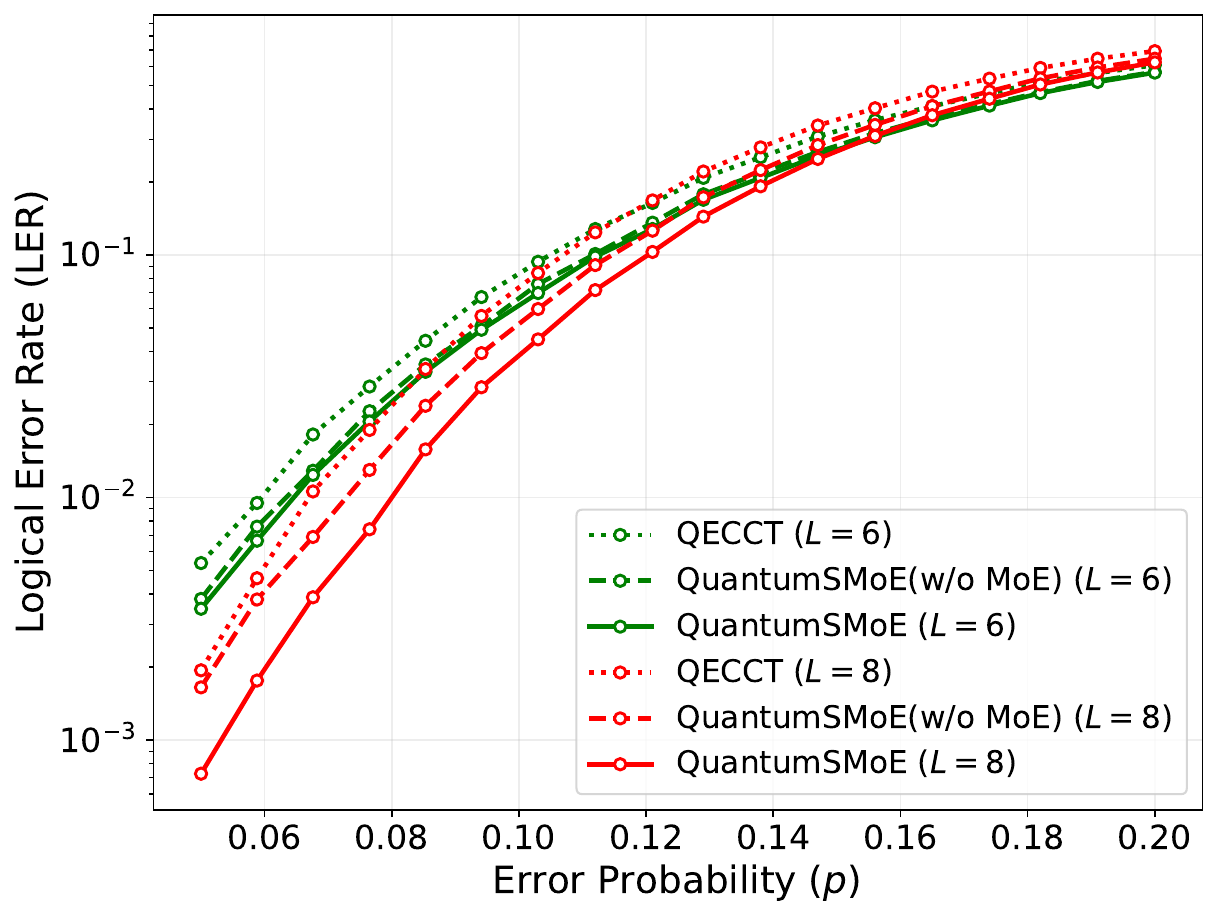}
    \caption{LER comparison  for the proposed QuantumSMoE with and without the MoE layer for distance $L=6,8$. QECCT is included as a baseline to demonstrate the effectiveness of patch embedding in extracting local topological information.}
    \label{fig:MoEorNOT}
\end{figure} 

Next, we examine the impact of ViT based architecture and MoE integration in our decoder. Figure~\ref{fig:MoEorNOT} evaluates the efficacy of the MoE block within our quantum decoder. Since our model utilizes a ViT backbone, it inherently incorporates local information through patch embeddings.  We further compare our approach with QECCT to demonstrate how the inclusion of patch embeddings provides a crucial spatial inductive bias. Unlike the standard Transformer framework of QECCT, which processes syndromes without explicit localized context, our model leverages these embeddings to effectively capture the underlying connectivity between qubits. Figure~\ref{fig:MoEorNOT} illustrates that even without the MoE component, QuantumSMoE consistently outperforms QECCT for distances $L=6$ and $8$. These findings suggest that incorporating spatial information is crucial for the model to learn the underlying correlations between errors. Notably, the addition of MoE layers provides a significant performance boost with very little overhead on computational cost. This integration enables the model to achieve decoding performance that consistently outperforms prior learning based decoders under the considered settings. 
\subsection{Ablation Study}
This section analyzes the effectiveness of the proposed QuantumSMoE model. We first demonstrate that the proposed PlusConv2D and Adaptive Masking mechanisms contribute to enhance the decoding accuracy compared to configurations without these components. Next, we evaluate the impact of the auxiliary loss function $\mathcal{L}_{\text{os}}$, highlighting its role in enhancing routing specialization and improving the overall performance of the SoftMoE block. 

\textbf{Impact of PlusConv2D and Adapting Mask.}
\begin{table}[]
\resizebox{\columnwidth}{!}{
\begin{tabular}{c|cccccc}
\midrule
\multirow{2}{*}{Model}                        & \multicolumn{6}{c}{Physical Error Rate (p)}         \\  
& $p=0.05$ & $p=0.07$ & $p=0.09$ & $p=0.11$ & $p=0.13$ & $p=0.15$\\ \midrule
QuantumSMoE (w/o PlusConv2D \& Adaptive Mask) & 0.0038   & 0.0129   & 0.0509   & 0.1010  & 0.1780  & 0.2660  \\ \midrule
QuantumSMoE (w/o Adaptive Mask)               & 0.0037   & 0.0128   & 0.0514   & 0.1020  & 0.1740  & 0.2620  \\ \midrule
QuantumSMoE (w/o PlusConv2D)                  & 0.0037   & 0.0127   & 0.0500   & 0.1000  & 0.1760  & 0.2650  \\ \midrule
QuantumSMoE                                   & $\mathbf{0.0035}$   & $\mathbf{0.0124}$   & $\mathbf{0.0492}$   & $\mathbf{0.0985}$   & $\mathbf{0.1690}$  & $\mathbf{0.2560}$  \\ \midrule
\end{tabular}
}
\caption{Impact of PlusConv2D and Adaptive Masking on decoding performance for the $L=6$ Toric code. The table presents the LER of QuantumSMoE and its ablated variants across various physical noise levels.} 
\label{table:plus_mask}
\end{table}
\begin{figure}
    \centering
    \includegraphics[width=0.5\linewidth]{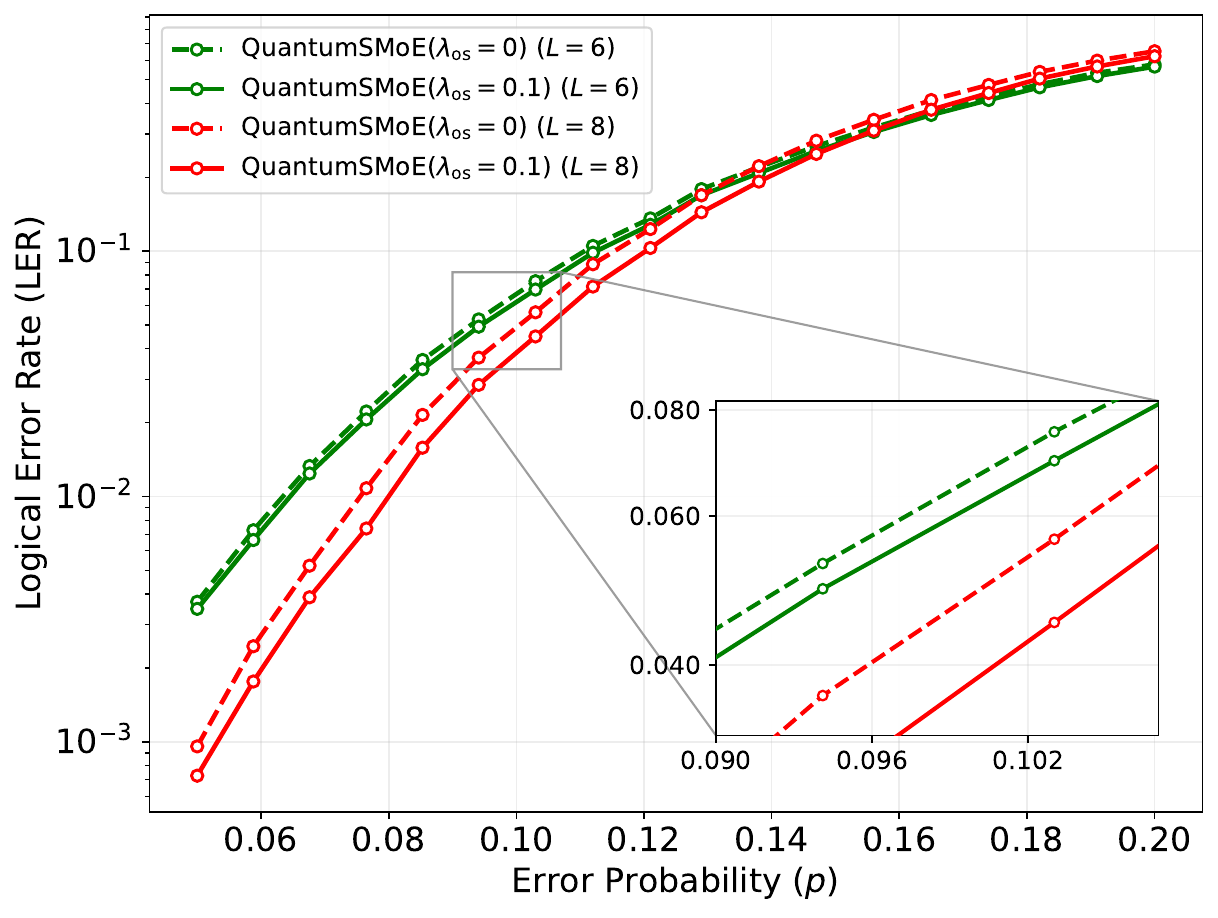}
    \caption{Effects of the slot orthogonal loss $\mathcal{L}_{\text{os}}$.}
    \label{fig:os_loss}
\end{figure}
Table~\ref{table:plus_mask} details the contributions of the proposed PlusConv2D layer and Adaptive Masking mechanism to the decoder's performance for the $L=6$ Toric code. To isolate the impact of each component, we conducted an ablation study by systematically removing them from the architecture. The results show a consistent reduction in Logical Error Rate (LER) across all benchmarks when these modules are integrated, with the performance gain becoming more pronounced as the physical error rate increases. This suggests that the local correlations between data qubits and their associated syndromes are effectively captured by our model.

\textbf{Impact of auxiliary loss function $\mathcal{L}_{\text{os}}$.} 
We further investigate the efficacy of the proposed slot-orthogonal loss function, $\mathcal{L}_{\text{os}}$, as illustrated in Figure~\ref{fig:os_loss} for distances $L=6$ and $L=8$. In both cases, the integration of $\mathcal{L}_{\text{os}}$ enables the model to converge toward a more accurate decoder. While the performance improvement is marginal at $L=6$, the gap becomes significantly more pronounced at $L=8$. These results suggest that this loss function is vital to the performance of SoftMoE and becomes increasingly critical as the surface length increases.

\textbf{The contribution of MoE Layer}.
\begin{figure}
    \centering
\centering
\subfloat[Syndrome pattern]{%
  \includegraphics[width=0.44\textwidth]{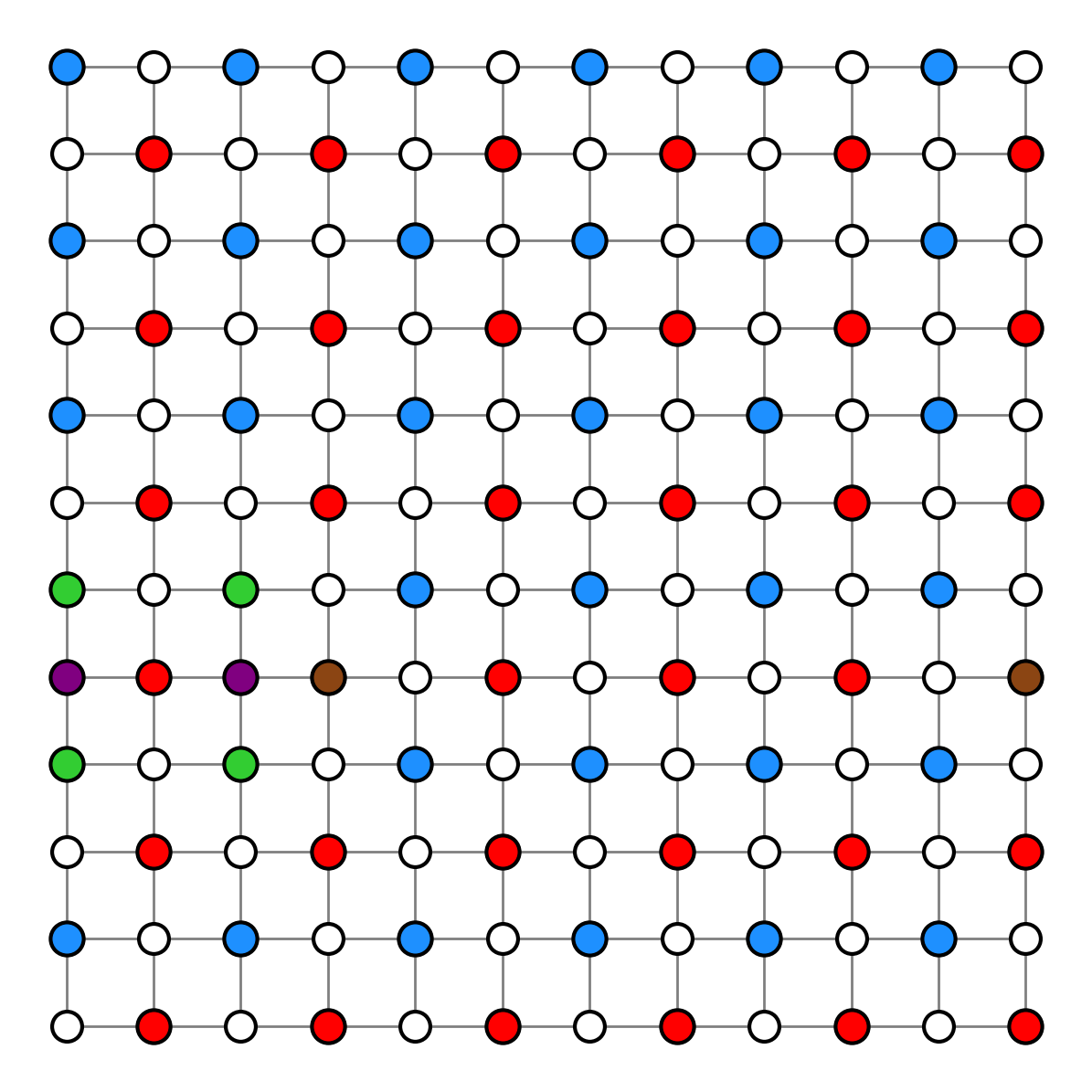}%
  \label{fig:LER_L6}%
}\hfill
\subfloat[Expert $7$, Slot $1$]{%
  \includegraphics[width=0.48\textwidth]{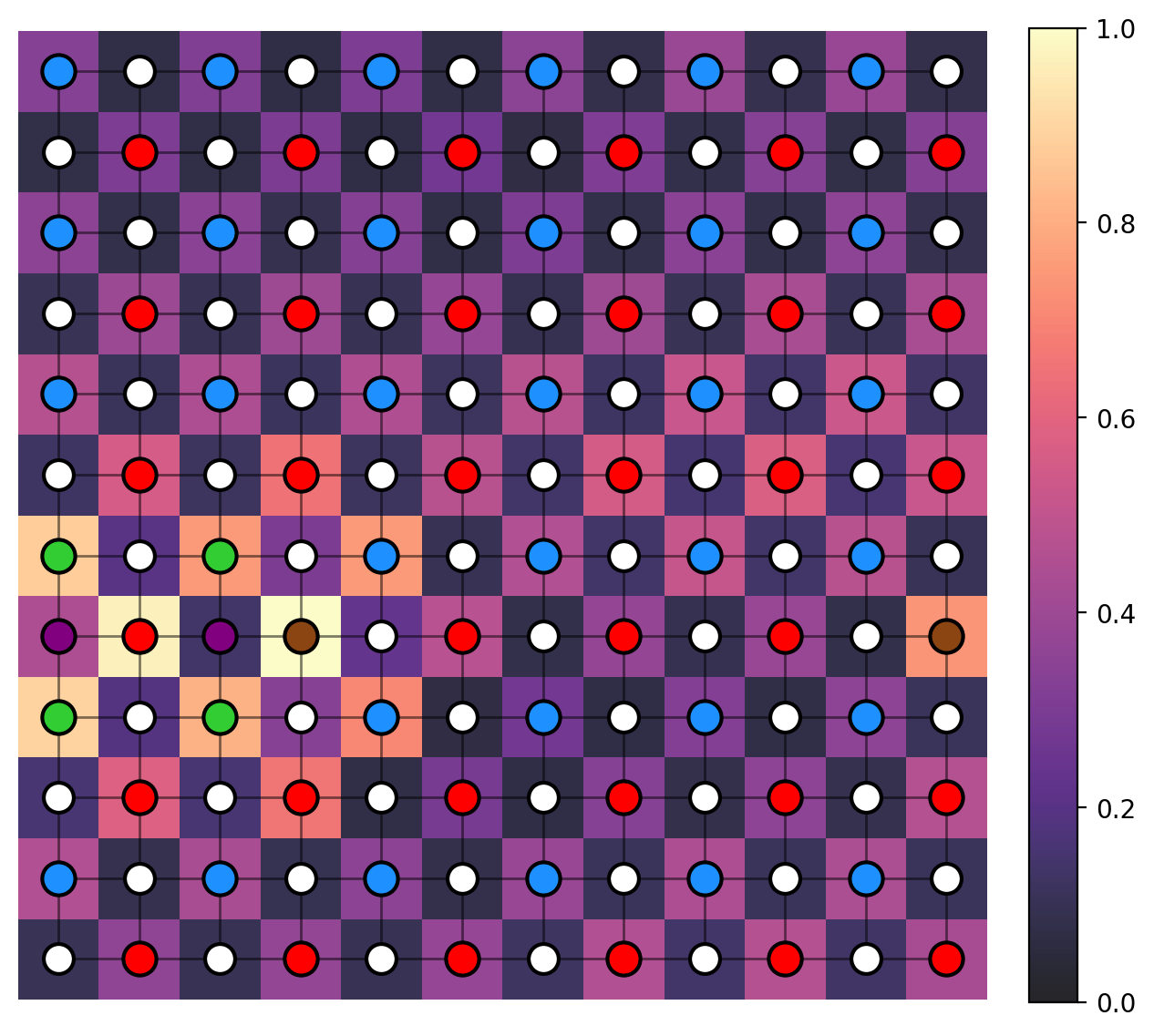}%
  \label{fig:LER_L8}%
}
    \caption{Visualization of slot linear combinations within the first SoftMoE layer (fifth block) using an $L=6$ syndrome instance. The selected slot is cherry-picked to demonstrate the efficacy of the slot assignment mechanism.  }
    \label{fig:moe_contribution}
\end{figure}
We further illustrate how incorporating an MoE layer into the decoder enhances interpretability of the decoding process. Figure~\ref{fig:moe_contribution} presents an experiment conducted on the toric code with $L=6$. As shown, the first slot of the seventh expert primarily attends to a specific error pattern within the syndrome. Consequently, the seventh expert is required to learn only a limited subset of error configurations, which is intractable for a conventional MLP to capture. By integrating a SoftMoE layer together with the slot orthogonality loss $\mathcal{L}_{\text{os}}$, the decoder is encouraged to isolate and emphasize critical syndrome information, thereby simplifying error learning for individual experts and providing deeper decoding insight. 

\section{Conclusion}
We present QuantumSMoE, a novel decoder that combines the spatial inductive bias of Vision Transformers with the scalability of mixture-of-experts architectures. Our framework uses plus-shaped embeddings and adaptive masking to better capture locality in topological stabilizer codes. Experiments on the toric code demonstrate significant reductions in logical error rate compared with established classical and ML-based baselines under depolarizing noise. Future work will extend this approach to higher-dimensional topological stabilizer codes and to more realistic noise models, including syndrome measurement noise and circuit noise.

\newpage
\phantomsection
\addcontentsline{toc}{section}{Bibliography}
\bibliographystyle{alpha}
\bibliography{bibliofile}
\newpage 
 \appendix
 
\end{document}